# Ferromagnetism induced by hybridization of Fe 3d orbitals with ligand InSb bands in n-type ferromagnetic semiconductor (In,Fe)Sb

Ryo Okano[1], Tomoki Hotta[1] Takahito Takeda[1], Kohsei Araki[1],

Kengo Takase[1], Le Duc Anh[1], Shoya Sakamoto[2], Yukiharu Takeda[3], Atsushi Fujimori[4,5],

Masaaki Tanaka[1,6], and Masaki Kobayashi[1,6,*]

[1]*Department of Electrical Engineering and Information Systems, The University of Tokyo, 7-3-1 Hongo, Bunkyo-ku, Tokyo 113-8656, Japan*

[2]*The Institute for Solid State Physics, The University of Tokyo, 5-1-5 Kashiwanoha, Kashiwa, Chiba 277-8581, Japan*

[3]*Materials Sciences Research Center, Japan Atomic Energy Agency, Sayo-gun, Hyogo 679-5148, Japan*

[4]*Department of Physics, The University of Tokyo, 7-3-1 Hongo, Bunkyo-ku, Tokyo 113-0033, Japan*

[5]*Department of Applied Physics, Waseda University, Okubo, Shinjuku, Tokyo 169-8555, Japan*

[6]*Center for Spintronics Research Network, The University of Tokyo, 7-3-1 Hongo, Bunkyo-ku, Tokyo 113-8656, Japan*

(Date: 8th Feb., 2022)

*Author to whom all correspondence should be addressed: masaki.kobayashi@ee.t.u-tokyo.ac.jp

## ABSTRACT

Fe-doped III-V ferromagnetic semiconductor (FMS) (In,Fe)Sb is a promising material for spintronic device applications because of the n-type carrier conduction and the ferromagnetism with high Curie temperature ($T_C > 300$ K). To clarify the mechanism of the high-$T_C$ ferromagnetism, we have investigated the electronic structure and magnetic properties of an $(In_{0.94},Fe_{0.06})Sb$ thin film by performing x-ray absorption spectroscopy (XAS) and x-ray magnetic circular dichroism (XMCD) measurements at the Fe $L_{2,3}$ edges. The magnetic-field ($\mu_0 H$) dependence of the XMCD spectra reveals that there are ferromagnetic-like Fe and paramagnetic-like Fe components in the (In,Fe)Sb thin film.

The XAS and XMCD spectra of the ferromagnetic-like and paramagnetic-like Fe components resemble those of other Fe-doped FMSs and extrinsic oxides, respectively. The finite value of the ratio between the orbital and spin magnetic moments estimated by applying the XMCD sum rules indicates that the valence state of the Fe ions substituting for the In sites in (In,Fe)Sb is not purely ionic $Fe^{3+}$, but intermediate between $Fe^{3+}$ and $Fe^{2+}$. The qualitative correspondence between the $\mu_0 H$ dependence of the visible-light magnetic circular dichroism intensity and that of the XMCD intensity demonstrates that the Zeeman splitting of the InSb band is proportional to the net magnetization of the doped Fe. These results suggest that the ferromagnetism of (In,Fe)Sb originates from the Fe $3d$ orbitals hybridized with the host InSb bands.

## I. INTRODUCTION

Ferromagnetic semiconductors (FMSs) exhibit both the properties of semiconductors and ferromagnets simultaneously and thus are promising materials for spintronics devices [1-3], which exploits both the charge and spin degrees of freedom of electrons. In III-V FMSs, magnetic elements such as Mn and Fe partially replace the group III sites. The Fe-doped III-V FMSs, (In,Fe)As [4-6], (In,Fe)Sb [7,8], (Ga,Fe)Sb [9-11] and (Al,Fe)Sb [12], have been successfully grown by molecular beam epitaxy (MBE) in the last decade. Since the doped Fe ions substitute for group III elements as $Fe^{3+}$, it has been believed that the Fe ions in a III-V semiconductor matrix do not simply act as donors or acceptors. In fact, various carrier types of Fe-doped III-V FMSs have been realized by co-doping or defect control: (In,Fe)As:Be [4] and (In,Fe)Sb [7] are n-type, (Ga,Fe)Sb [9] is p-type, and (Al,Fe)Sb [12] is insulating. Moreover, (Ga,Fe)Sb and (In,Fe)Sb show ferromagnetism whose Curie temperature ($T_C$) is higher than room temperature. The highest $T_C$ of $(Ga_{1-x},Fe_x)Sb$ is about 400 K at Fe concentration $x = 0.20$ [13], and that of $(In_{1-x},Fe_x)Sb$ is about 385K at $x = 0.35$ [14]. Therefore, Fe-doped III-V FMSs are promising materials for the realization of spintronics devices with pn junctions [15] operating at room temperature.

Various studies of the physical properties of n-type (In,Fe)Sb have been carried out to clarify the origin of the ferromagnetism [7,8,16]. X-ray diffraction and scanning transmission electron microscopy measurements indicated that $(In_{1-x},Fe_x)Sb$ maintains the zinc-blende-type crystal structure up to at least $x \leqq 0.16$ [7]. Magnetic circular dichroism (MCD) in a visible-light range ($1 - 5$ eV) and anomalous Hall effect measurements have confirmed intrinsic ferromagnetism in (In,Fe)Sb, in which ferromagnetic order appears in the zinc-blende semiconductor phase (there is no visible evidence for secondary phases) [7]. A first-principles calculation [16] for (In,Fe)Sb has predicted that the isoelectronic Fe dopants induce antiferromagnetic interaction between the Fe ions through the super-exchange mechanism and the transition from the antiferromagnetic to ferromagnetic states is induced by additional carrier doping. This behavior can be well understood in terms of the Alexander-Anderson-Moriya mechanism [17,18]. The electrical control of ferromagnetism in (In,Fe)Sb by applying a gate voltage indicated that both the electron-carrier-induced ferromagnetic interaction and the super-exchange mechanism contribute to the emergence of the ferromagnetism in (In,Fe)Sb [8].

For further understanding of the origin of magnetism in (In,Fe)Sb, it is necessary to characterize and reveal the electronic states of the doped Fe ions related to the ferromagnetism in detail. To address this issue, we investigate the relationship between the local electronic states of the Fe ions in (In,Fe)Sb and the ferromagnetic behavior by x-ray absorption spectroscopy (XAS) and x-ray magnetic circular dichroism (XMCD). Synchrotron radiation based XMCD is an element-specific magnetic probe and a powerful tool to study the electronic state of the doped magnetic ions in FMSs [19-23]. The experimental findings based on the XMCD measurements suggest that the ferromagnetism in (In,Fe)Sb is intrinsic and originates from the Fe $3d$ orbitals hybridized with the ligand bands of the host InSb.

## II. EXPERIMENTAL

An $(In_{0.94},Fe_{0.06})Sb$ thin film with a thickness of 15 nm was grown on a p-type GaAs(001) substrate by MBE. In order to avoid surface oxidation, the sample was covered with a thin amorphous As capping layer after the MBE growth of the $(In_{0.94},Fe_{0.06})Sb$ layer. The sample structure is, from top to bottom, As capping layer ~1 nm/$(In_{0.94},Fe_{0.06})Sb$ 15 nm/AlSb 100 nm/AlAs 6 nm/GaAs:Be 100 nm grown on a $p^+$ GaAs (001) substrate. The $T_C$ of the sample was about 100 K estimated by the Arrott plot of visible-light MCD intensity - perpendicular magnetic field ($\mu_0H$) characteristics.

XAS and XMCD measurements were performed at beamline BL23-SU of SPring-8. The measurements were conducted under an ultrahigh vacuum below $1.0\times10^{-8}$ Pa at a temperature ($T$) of 10 K. Circularly polarized x-rays in the energy range of 690 – 740 eV near the Fe $L_{2,3}$ absorption edges were used for the measurements. The absorption spectra for circularly polarized x-rays were obtained by reversing photon helicity at each photon energy ($h\nu$ or $\hbar\omega$) and were taken in the total electron yield mode. Here, XAS spectra taken with left and right circularly polarized x-rays are defined as $\mu^+$ and $\mu^-$, respectively, and then the XAS spectrum $\mu(\omega)$ and the XMCD spectrum $\Delta\mu_H(\omega)$ are represented as $\mu(\omega) = (\mu^+ + \mu^-)/2$ and $\Delta\mu(\omega) = (\mu^+ - \mu^-)$, where $\hbar\omega$ is the photon energy. Magnetic fields were varied from -7 T to 7 T and applied parallel to the incident x-rays corresponding to the surface-normal direction. The sample was divided into two pieces and one of them was etched by HCl to obtain the clean surface. We etched the sample with HCl (2.4 mol/L) for 5 s to remove the capping layer and subsequently rinsed it with water just before loading the sample in the vacuum chamber of the spectrometer, following the same procedure reported elsewhere [19].

## III. RESULTS AND DISCUSSION

Figure 1(a) shows XAS spectra of the as-grown and HCl-etched $(In_{0.94},Fe_{0.06})Sb$ thin films at the Fe $L_{2,3}$ absorption edges. The XAS spectrum of the as-grown sample shows two peaks at $hv \sim 707.7$ eV and $\sim 709.7$ eV in the Fe $L_3$ edge, while the XAS spectrum of the HCl-etched sample shows a single peak at $hv \sim 707.7$ eV. This spectral line-shape difference indicates that there are two Fe components in the film: the component having a peak at 709.7 eV, which disappears by the HCl etching, and the component having a peak at 707.7 eV, which remains after the HCl etching.

Figure 1(b) shows the XMCD spectra of the as-grown film with varying $\mu_0 H$. At the Fe $L_3$ edge, only the peak around 707.7 eV is observed in the XMCD spectrum taken with $\mu_0 H = 0.1$ T, while the peak around 709.7 eV increases with increasing $\mu_0 H$. This suggests that there are mainly two kinds of magnetic components in the $(In_{0.94},Fe_{0.06})Sb$ thin film, and the components having the peaks at 707.7 eV and 709.7 eV are ferromagnetic-like (FM-like) and paramagnetic-like (PM-like), respectively. The peak positions of the two peaks in the XMCD spectra are almost the same as those in the XAS spectra, as shown by the vertical dashed lines in Fig. 1. The XAS and XMCD peaks around 707.7 eV are also observed in other Fe-doped FMSs, such as (Ga,Fe)Sb [19] and (In,Fe)As:Be [22], while the peak position at $\sim$ 709.7 eV coincides with that of $\gamma$-Fe$_2$O$_3$ [24].

Figure 2 shows the $\mu_0 H$ dependence of the XMCD intensities (XMCD - $H$ curve) measured at $h\nu$ = 707.7 eV and 709.7 eV. The XMCD - $H$ curve measured at 707.7 eV steeply increases near the zero magnetic field (0 T < $\mu_0 H$ < 1 T) and gradually increases above $\mu_0 H$ ~1 T. In contrast, the steep increase is almost absent for the XMCD - $H$ curve measured at 709.7 eV, but the linear gradual increase is dominant. These results indicate that the two components having peaks at 707.7 eV and 709.7 eV in the measured thin film are predominantly FM and PM, respectively.

Since the $\mu_0 H$ dependence of the XMCD intensities indicates that there are two components in (In,Fe)Sb, it is necessary to clarify the origin of each component. We denote the FM-like and PM-like components by $\alpha$ and $\beta$. The XMCD spectrum and the XAS spectrum were decomposed into the two components respectively by the following procedure: Firstly, assuming that the magnetization of the $\beta$ component responds linearly to the magnetic field, the XMCD spectra of the $\alpha$ and $\beta$ components ($\Delta\mu_\alpha$ and $\Delta\mu_\beta$) were obtained by the following equations:

$$\Delta\mu_\beta = \frac{7}{3}(\Delta\mu_{7\text{T}} - \Delta\mu_{4\text{T}}), \qquad (1)$$

$$\Delta\mu_\alpha = \Delta\mu_{7\text{T}} - \Delta\mu_\beta. \qquad (2)$$

Here, $\Delta\mu_{n\text{T}}$ is the XMCD spectrum at $\mu_0 H$ = n T. Secondly, we conducted the decomposition of the XAS spectra. Since XAS spectral line shapes are usually insensitive to $\mu_0 H$, the XAS spectra cannot be decomposed from the $\mu_0 H$ dependence as done for the XMCD spectra. Therefore, we have extracted those Fe components by comparing the XAS spectra before and after the HCl etching as in the previously reported XAS spectra of (Ga,Fe)Sb [19]. Since the peak position of 707.7 eV in the XAS spectrum after the HCl etching corresponds to that of the FM-like component in the XMCD spectra, it is likely that the α component is predominant in the XAS spectrum after the HCl etching. Therefore, we have adopted the XAS spectrum after the HCl etching as the XAS spectrum for the α component ($\mu_\alpha$). The XAS spectrum for the β component ($\mu_\beta$) is then obtained by subtracting a fraction of $\mu_\alpha$ spectrum from the XAS spectrum of the as-grown sample such that the α-component shoulder at 707.7 eV disappears in the $\mu_\beta$ spectrum.

Figure 3 shows the XAS and XMCD spectra decomposed into the α and β components at $\mu_0 H$ = 7 T. The XAS spectrum of the α component shows a peak only at 707.7 eV in the Fe $L_3$ edge. The spectral line shapes of the XAS and XMCD spectra of the α component resemble those of other FMSs such as (Ga,Fe)Sb [19], (In,Fe)As [22], (Al,Fe)Sb [20]. This indicates that the electronic state of the α component is close to those of the substitutional Fe ions in the other Fe-doped FMSs and that the α component is intrinsic to (In,Fe)Sb. On the other hand, the spectral line shapes of the XAS and XMCD spectra of the β component having peaks at 708 eV and 709.7 eV in the Fe $L_3$ edge are similar to those of $\gamma$-Fe$_2$O$_3$ [24]. Since the HCl etching is considered to remove the layers near the surface, the β component removed by the HCl etching is most likely an extrinsic component such as surface oxides.

To identify the electronic states of the α and β components in more detail, we have estimated the magnetic moments using the XMCD sum rules [25,26]. By applying the XMCD sum rules to the obtained XMCD spectra $\Delta\mu(\omega)$, the spin and orbital magnetic moments of the doped Fe ions in units of $\mu_B$/atom are estimated separately. The XMCD sum rules are as follows:

$$M_{orb} = -\frac{2q}{3r}(10 - N_d),\qquad(3)$$

$$M_{spin} \approx -\frac{3p - 2q}{r}(10 - N_d).\qquad(4)$$

Here, $p = \int_{L_3} \Delta\mu(\omega)d\omega$, $q = \int_{L_{2,3}} \Delta\mu(\omega)d\omega$, $r = \int_{L_{2,3}} \mu(\omega)d\omega$ and $N_d$ is the number of $3d$ electrons. It should be noted here that the ratio between $M_{orb}$ and $M_{spin}$ can be estimated from the XMCD spectra without the values of $r$ and $N_d$. The estimated values of $M_{orb}/M_{spin}$ for the α and β components at $\mu_0H = 7$ T and $T = 10$ K are listed in Table I. The $M_{orb}/M_{spin}$ value of the α component, which appears to be intrinsic to (In,Fe)Sb, has a finite positive value of 0.06 ± 0.01. This value is close to 0.065 ± 0.014 reported for $(In_{0.95},Fe_{0.05})As:Be$ [21] and larger than 0.043 ± 0.001 for Fe (bcc) [27]. Therefore, the electronic state of the substitutional Fe ions in (In,Fe)Sb is similar to that of (In,Fe)As:Be but is different from that of Fe (bcc). Since the orbital magnetic moment of an ionic $Fe^{3+}$ ion ($d^5$ high-spin state) is zero, the finite $M_{orb}/M_{spin}$ suggests that the valence of Fe in (In,Fe)Sb is different from the purely ionic $Fe^{3+}$. The valence state of the doped Fe ions is likely an intermediate state between trivalent and divalent due to a charge transfer from the ligand Sb to the Fe $3d$ orbitals via the hybridization with the surrounding ligand Sb bands. Assuming that $C_1|d^5> + C_2|d^6\underline{L}>$ as the Fe $3d$ state, which is a superposition of the states $d^5$ and $d^6\underline{L}$ ($\underline{L}$ is the hole in the ligand), the values of $C_1^2$ and $C_2^2$ for the α component are estimated to be 0.94 and 0.06, respectively, from $M_{orb}/M_{spin}$. Therefore, the number of electrons $N_d$ is estimated to be 5.06. The values of $M_{orb}$ and $M_{spin}$ estimated by applying $N_d$ = 5.06 are shown in Table I. Here, the applied correction factor is 0.685 ($d^5$) [28]. Since the $M_{orb}/M_{spin}$ value for the β component is almost zero, the β component probably originates from trivalent $Fe^{3+}$ ($d^5$) oxides like $Fe_2O_3$.

To further clarify the magnetic behavior, the $\mu_0H$ dependences of the magnetizations ($M$ - $H$ curves) are studied. The inset of Fig. 3(b) shows the $M$ - $H$ curves. Here, the magnetization values estimated from the XMCD sum rules are in units of $\mu_B$ per Fe atom.

While the magnetization of the β component is linearly proportional to $\mu_0 H$, the magnetization of the α component shows FM-like behavior with $\mu_0 H$s. It should be noted here that the magnetization of the α component is not saturated even at the highest $\mu_0 H$, and the magnetization at 7 T of ~2.8 $\mu_B$/Fe is smaller than the full moment of 5 $\mu_B$/Fe for $Fe^{3+}$ ($d^5$). This suggests that the magnetic behavior of the doped Fe ions in (In,Fe)Sb involves not only a FM component, but also a PM component. The details of the magnetic components are discussed later.

Figure 4(a) shows comparison between the $M$ - $H$ curve of the intrinsic Fe component (α) and the MCD - $H$ curve from visible-light MCD measurements. Since the MCD is the difference between the intensities of the transition from the valence band to the conduction band with circularly polarized lights under an applied $\mu_0 H$, the MCD signal reflects the Zeeman splitting of the host InSb bands. As shown in Fig. 4(a), the MCD - $H$ curve is almost identical to the $M$ - $H$ curve estimated from the XMCD spectra, indicating that the magnetic behavior of the doped Fe ions in (In,Fe)Sb is proportional to the Zeeman splitting of the InSb bands. It should be mentioned here that the net magnetization of the intrinsic Fe component contributes to the magnitude of the Zeeman splitting, although the intrinsic Fe of (In,Fe)Sb contains not only a FM component but also a PM component. Considering the finite $M_{orb}$ discussed above, the substitutional Fe ions in (In,Fe)Sb induce the Zeeman splitting of the host InSb via the hybridization between the Fe 3$d$ orbital with the ligand $sp$ bands. Based on these findings, we conclude that the ferromagnetism in (In,Fe)Sb originates from the hybridization between the Fe 3$d$ orbitals and the valence or conduction bands of InSb, but not from oxidized Fe or precipitated Fe metal.

As we noted above, the intrinsic component involves both the FM-like and PM-like components. To further investigate this magnetization process of the intrinsic component, we have analyzed the shape of the $M$ - $H$ curve. Note that the local electronic state of the doped Fe ions is common for these FM-like and PM-like components within the experimental accuracy. This suggests that the different magnetic behavior of the doped Fe ions comes from the inhomogeneous distribution of the Fe ions in (In,Fe)Sb [20,22]. While the PM component likely originates from isolated Fe ions in regions with low Fe density, regions with high Fe density may contribute to the FM and/or superparamagnetic (SPM) magnetic behavior depending on the size of the regions (Fe domain). In order to determine the ratio of the FM, SPM, and PM components, the $M$ - $H$ curve has been fitted using the following functions:

$$M = s m_{\text{sat}} L\left(\frac{\mu_{\text{FM}}\mu_0 H}{k_{\text{B}}T}\right) + t m_{\text{sat}} L\left(\frac{\mu_{\text{SPM}}\mu_0 H}{k_{\text{B}}T}\right) + (1 - s - t)\frac{C\mu_0 H}{T + T_A} \ . \qquad (5)$$

Here, the FM and the SPM components are assumed to be represented by Langevin functions, and the PM component is represented by a linear function. $s$ ($t$) is the ratio of Fe atoms participating in the FM (SPM), $m_{\text{sat}}$ is the total magnetic moment of the Fe atom, $\mu_{\text{FM}}$ ($\mu_{\text{SPM}}$) is the magnitude of the magnetic moment per FM (SPM) region where the magnetic moments are aligned, $C$ is the Curie constant, $T_A$ is the Weiss temperature, and $k_{\text{B}}$ is the Boltzmann constant. We have assumed that $m_{\text{sat}} = 5\ \mu_{\text{B}}$, which is the total magnetic moment of $Fe^{3+}$. As shown in Fig. 4(b), the fitting well reproduces the $M$ - $H$ curve, suggesting that all the three magnetic components contribute to the intrinsic magnetic behavior. Table II lists the fitted results of the parameters. The percentages of FM, SPM, and PM are about 27 %, 13 %, and 60 %, respectively. Note that we have also tried fitting with a single Langevin function and a linear function, but the $M$ - $H$ curve obtained by the experiment is not reproduced well. $T_A$ is about 28 K, which is close to the Weiss temperature of PM $(Ga_{0.96},Fe_{0.04})As$ without carrier doping (32 K) [29]. This implies that the magnetic interaction between the isolated PM Fe ion in (In,Fe)Sb is antiferromagnetic, consistent with previous theoretical calculations [16]. The magnetic moment per FM domain $\mu_{\text{FM}}$ is deduced to be about 840 $\mu_{\text{B}}$ – 1025 $\mu_{\text{B}}$, which corresponds to 168 – 205 Fe atoms in each FM domain on average, and $\mu_{\text{SPM}}$ is about 80 $\mu_{\text{B}}$ – 125 $\mu_{\text{B}}$, which corresponds to 16 – 25 Fe atoms in each SPM domain on average. The formation of Fe-rich domains is consistent with the attractive interaction of Fe atoms in (In,Fe)Sb at the second nearest neighbor site, as predicted by theoretical calculations [16]. Although the majority of the Fe ions seems isolated in (In,Fe)Sb and shows PM behavior, a significant fraction of the Fe ions are located close to each other and exhibit the FM or SPM behavior. As the Fe doping concentration is increased, the fraction of Fe atoms involved in the FM region is expected to increase. This model is consistent with the experimental results that $T_C$ increases with increasing Fe doping concentration [7,14]. To elucidate the magnetic behavior of (In,Fe)Sb in more detail, the XMCD studies on (In,Fe)Sb with different Fe concentrations are desirable.

## IV. CONCLUSION

In conclusion, we have performed XMCD measurements on an $(In_{0.94},Fe_{0.06})Sb$ thin film to investigate the electronic states of the doped Fe ions related to the ferromagnetism. The XAS and XMCD spectra taken at the Fe $L_{2,3}$ edges have been decomposed into the substitutional Fe ions of (In,Fe)Sb and extrinsic Fe oxides formed near the surface. The finite ratio of the orbital to spin magnetic moments estimated by applying the XMCD sum rules indicates that the valence state of the substitutional Fe ions in (In,Fe)Sb is not purely ionic $Fe^{3+}$, but intermediate between $Fe^{3+}$ and $Fe^{2+}$. The qualitative correspondence between the visible-light MCD - $H$ and XMCD - $H$ curves demonstrates that the Zeeman splitting of the InSb band is proportional to the net magnetization of the doped Fe atoms.

Based on these findings, we conclude that the ferromagnetism in (In,Fe)Sb originates from the Fe 3$d$ orbitals hybridized with the host InSb bands. In addition, the fitting result of the $M$ - $H$ curve suggests that the magnetism of (In,Fe)Sb consists of FM, PM, and SPM components, and that all of the magnetic components are derived from the orbital hybridization of the Fe 3$d$ orbitals with the host InSb bands. These results are important for understanding the physical properties of (In,Fe)Sb, such as the mechanism of ferromagnetism, and will be useful for applications such as spintronics devices using (In,Fe)Sb in the future.

## ACKNOWLEDGEMENT

This work was supported by a Grants-in-Aid (Nos. 20H05650, 18H05345), CREST (No. JPMJCR1777), and PRESTO Programs (Grant No. JPMJPR19LB) of the Japan Science and Technology Agency. This work was partially supported the Spintronics Research Network of Japan (Spin-RNJ). This work was performed under the Shared Use Program of Japan Atomic Energy Agency (JAEA) Facilities (Proposal No. 2018B-E23, 2020A-E18 and 2021A-E24) supported by JAEA Advanced Characterization Nanotechnology Platform as a program of "Nanotechnology Platform" of the Ministry of Education, Culture, Sports, Science and Technology (MEXT) (Proposal No. A-18-AE-

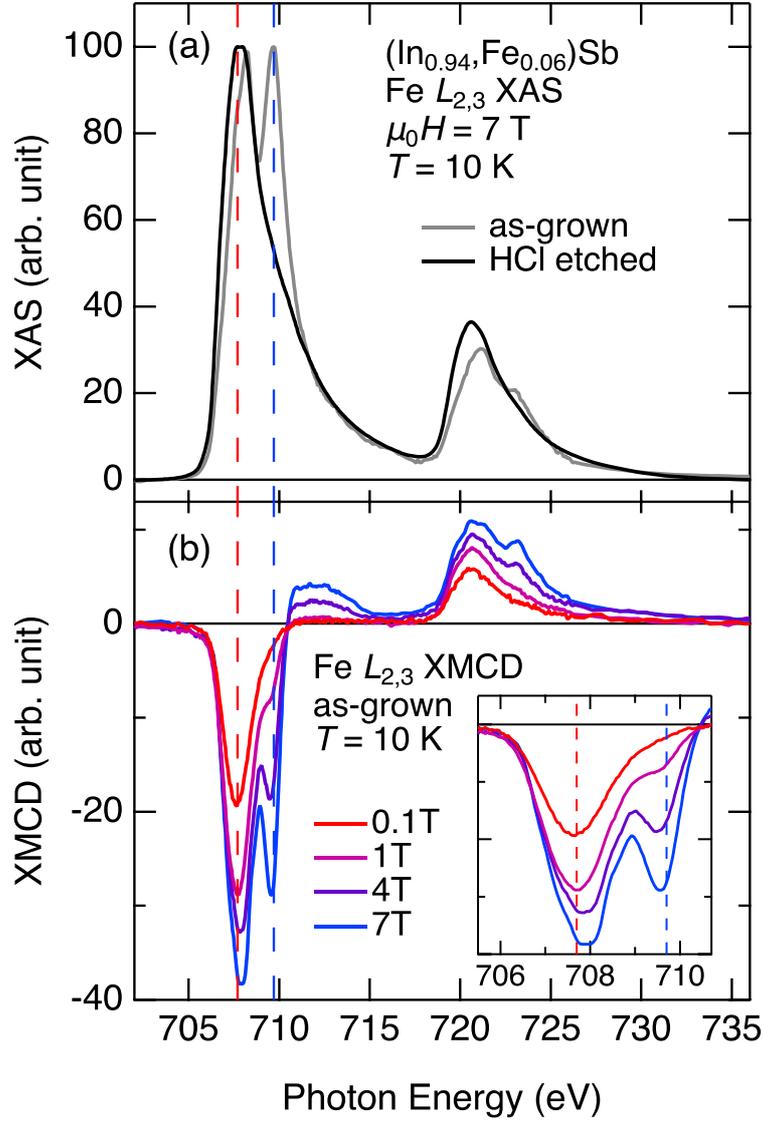

FIG. 1. XMCD spectra of (In$_{0.94}$,Fe$_{0.06}$)Sb thin films at Fe $L_{2,3}$ absorption edges. (a) XAS

spectra of the as-grown and HCl etched thin films. (b) XMCD spectra with varying $\mu_0 H$

of the as-grown film. These spectra are normalized to the maximum intensity of the XAS

spectrum as 100. The red dashed line is the peak position at $\mu_0 H$ = 0.1 T near 707.7eV.

The blue dashed line is the energy of the XMCD peak position at $\mu_0 H$ = 7 T near 709.7

eV. The inset shows an enlarged plot of XMCD spectra at Fe $L_3$ absorption edge.

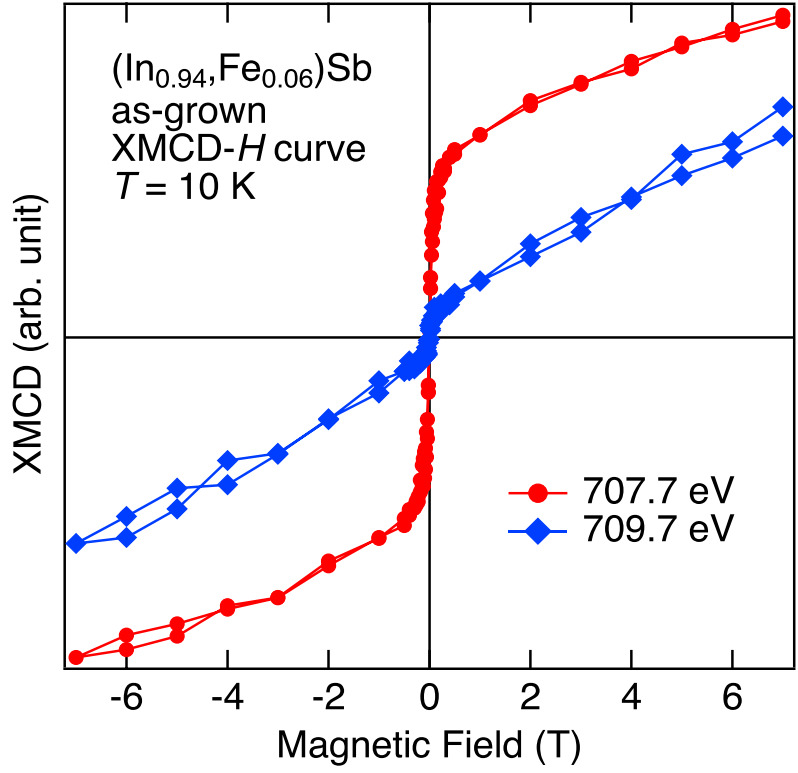

FIG. 2. Magnetic-field dependence of the XMCD intensity (XMCD - $H$ curve) of the as-grown $(In_{0.94},Fe_{0.06})Sb$ thin film. The red circle and blue rhombic markers are XMCD - $H$ curves taken at $h\nu \sim 707.7$ eV and $h\nu \sim 709.7$ eV, respectively.

TABLE I. Spin and orbital magnetic moments of the α and β in the $(In_{0.94}Fe_{0.06})Sb$ thin film at $\mu_0 H = 7$ T and $T = 10$ K. Here, the number of $d$ electrons is assumed to be 5.06. The correction factor for the Fe $3d$ ion (0.685) is employed [28].

|   | $M_{orb}/M_{spin}$ | $M_{orb}$ ($\mu_B$/Fe) | $M_{spin}$ ($\mu_B$/Fe) |
|---|---|---|---|
| α | $0.06 \pm 0.01$ | 0.16 | 2.60 |
| β | $0.016 \pm 0.003$ | 0.044 | 2.75 |

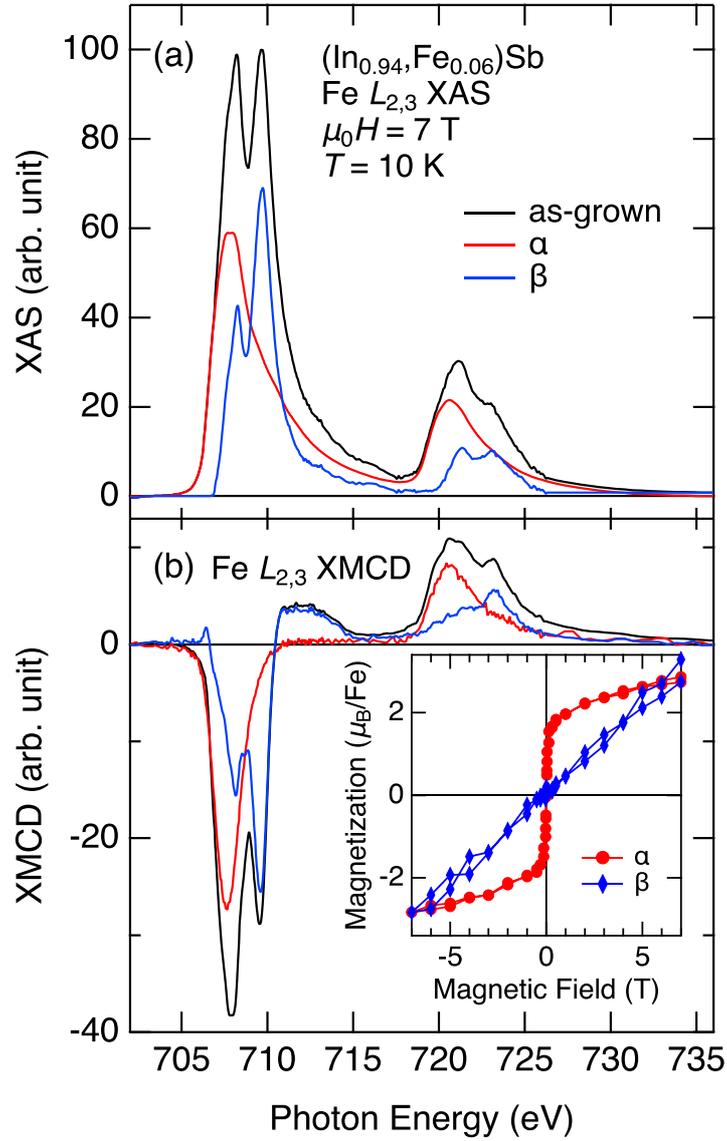

FIG. 3. Decomposition analysis for XAS and XMCD spectra of the as-grown $(In_{0.94},Fe_{0.06})Sb$ thin film at Fe $L_{2,3}$ edges. (a) XAS spectra and (b) XMCD spectra at $\mu_0 H$ = 7 T. The black line is the raw XMCD spectrum. The inset is the magnetic field dependence of the magnetization ($M$ - $H$ curve). Based on the XMCD sum rules [25,26], the vertical axis was converted from the XMCD intensity to the magnetization per Fe atom. The circle and rhombic markers are the α and β components, respectively.

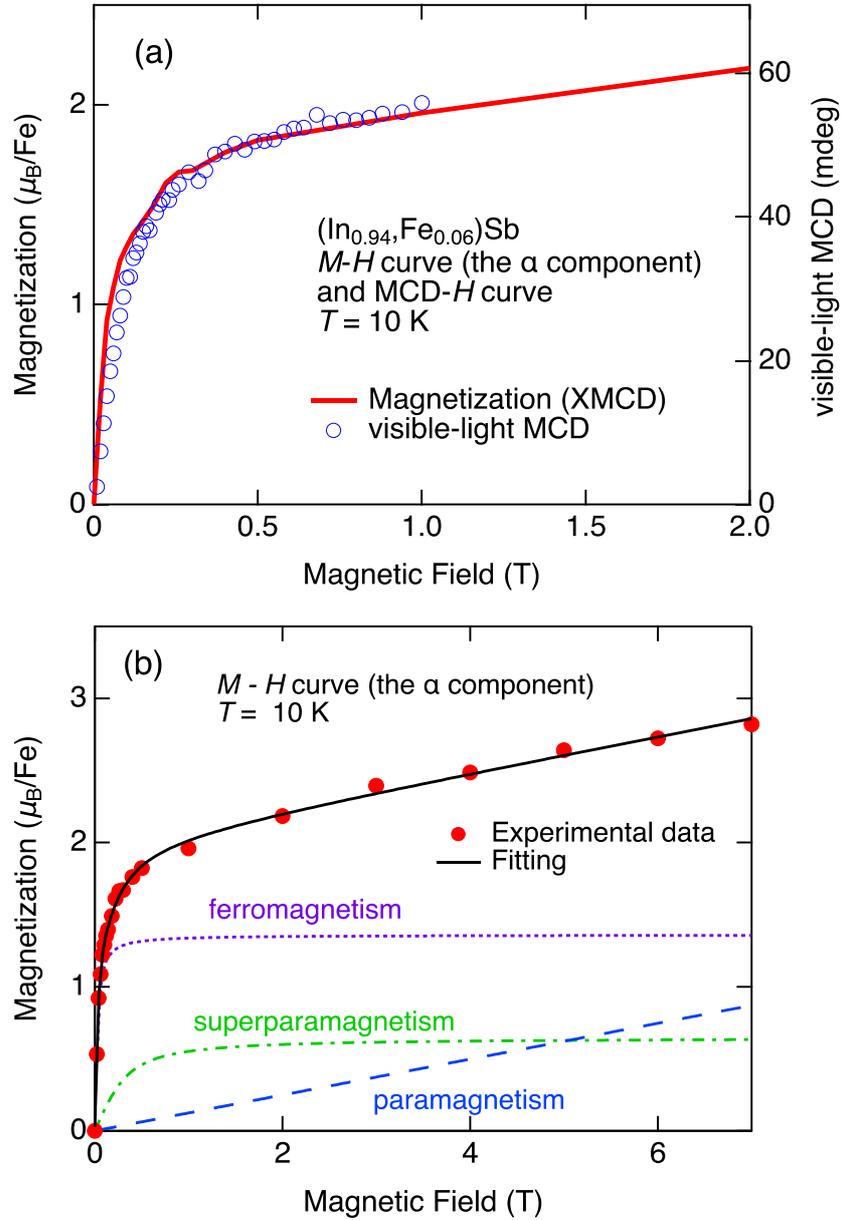

FIG. 4. *M - H* curves of the doped Fe ions in (In,Fe)Sb. (a) *M - H* curve of the α component and visible-light MCD - *H* curve of the (In$_{0.94}$,Fe$_{0.06}$)Sb thin films. (b) *M - H* curve of the α component and the fitting result by Langevin functions and liner function. Dotted, dash-dotted, and dashed lines are the FM, SPM and PM components, respectively. Solid line is the fitting result corresponding to the sum of all the components.

TABLE II. Fitting parameters using Eq. (5) for the $M$ - $H$ curve.

| $s$ | $\mu_{\text{FM}}$ | $t$ | $\mu_{\text{SPM}}$ | $T_A$ |
|---|---|---|---|---|
| $0.272 \pm 0.017$ | $933 \pm 92$ | $0.129 \pm 0.016$ | $102 \pm 22$ | $27.7 \pm 1.7$ |